\documentclass[%
 reprint,
superscriptaddress,
 amsmath,amssymb,
aps,
prl,
]{revtex4-2}

\usepackage{graphicx}
\usepackage{dcolumn}
\usepackage{bm}
\usepackage{color}

\newcommand\LLNL{Quantum Simulations Group, Lawrence Livermore National Laboratory, Livermore CA 94551, USA}

\begin{document}

\preprint{APS/123-QED}

\title{Dynamics of ballistic photocurrents driven by Coulomb scattering in a two-dimensional material}
\author{Liang Z. Tan}
\affiliation{Molecular Foundry, Lawrence Berkeley National Laboratory}
\author{Xavier Andrade}
\affiliation{\LLNL}
\author{Sangeeta Rajpurohit}
\affiliation{Molecular Foundry, Lawrence Berkeley National Laboratory}
\author{Alfredo A. Correa}
\affiliation{\LLNL}
\author{Tadashi Ogitsu}
\affiliation{\LLNL}

\begin{abstract}
First principles real-time time dependent density functional theory (rt-TDDFT) calculations reveal the existence of ballistic photocurrents generated by Coulomb scattering, {which} has not previously been considered as a mechanism for the bulk photovoltaic effect. With monolayer GeS as an example, it is predicted that ballistic currents can be comparable to shift currents under experimentally accessible conditions. 
 
\end{abstract}

\maketitle

The bulk photovoltaic effect (BPVE) is the generation of a direct current in an extended material under photoexcitation, independent of interfacial effects~\cite{dai-recent-2023}. 
Following early work to understand photocurrents in ferroelectrics~\cite{von-baltz-theory-1981}, this field~\cite{belinicher-photogalvanic-1980} has recently experienced a revival, driven by advances in first-principles computation~\cite{young-first-2012}, and the recasting of its response theory to the modern language of wavefunction geometry~\cite{morimoto-topological-2016,morimoto-geometric-2023,ahn-riemannian-2022,alexandradinata-topological-2022}. { The BPVE has attracted theoretical interest as a novel photocurrent mechanism as its influencing factors are not yet fully understood~\cite{tan-upper-2019}}. It can also be used as a powerful mode of light-matter control or detection via its sensitive dependence to light polarization and frequency.  

The shift current has been identified as one of the major contributions to the BPVE in polar materials~\cite{young-first-2012}. It is current generated directly by photoexcited transitions between valence and conduction bands. Much research on the BPVE operates within the assumption that the BPVE is dominated by the shift current~\cite{tan-shift-2016}. Within perturbation theory, the shift current of a perfect crystal of non-interacting electrons appears to have no dependence on carrier scattering effects. However, recent work has revealed that shift currently is strongly modified by scattering once these stringent conditions are removed~\cite{barik-nonequilibrium-2020, matsyshyn-rabi-2021}, with consequences for real materials with a number of intrinsic and extrinsic scattering sources.  

The importance of scattering is further highlighted when considering BPVE arising from relaxation and recombination dynamics~\cite{sturman-ballistic-2019, zhu-anomalous-2023}. Of these effects, ballistic photocurrents stand out for being driven entirely by scattering~\cite{kral-quantum-2000, menshenin-phonon-2003, deyo-semiclassical-2009, mahon-quantum-2019, okamura-photovoltaic-2022}. Here, carrier imbalance is established by scattering processes that break inversion symmetry. Theories for ballistic current driven by electron-phonon interactions~\cite{dai-phonon-assisted-2021, rajpurohit-ballistic-2023}  suggest that ballistic photocurrents may account for a significant portion of the total BPVE in materials where electron-phonon scattering is strong. Photo-Hall measurements have provided experimental support for the existence of ballistic photocurrents~\cite{burger-direct-2019}. On the other hand, previous theories for ballistic current driven by electron-hole scattering have not predicted strong ballistic currents, although only under weak-field assumptions inherent to perturbation theory~\cite{dai-first-principles-2021}.  

In this paper, we re-examine BPVE from the perspective of real-time TDDFT~\cite{runge-tddft-1984}, free from the assumptions of weak-field or weak-scattering. We find that ballistic photocurrents driven by Coulomb scattering are a major component of BPVE, being comparable to the shift current under high electric field strengths of 0.1 V/nm in 2D materials. By Coulomb scattering, we refer to the finite lifetime acquired by electrons  and holes as they interact with the electronic charge density.  As this form of scattering is intrinsic to any material, our results are likely to generalize to a wide range of materials. 
Previous real-time simulations~\cite{miyamoto-visualizing-2010, chan-giant-2021, he-ultrafast-2022, rajpurohit-nonperturbative-2022, rajpurohit-ballistic-2023} of photocurrent and time-resolved experiments~\cite{priyadarshi2012} have been useful for understanding transient photocurrent response, due to their nonperturbative and non-equilibrium nature. However, the explicit identification of the Coulomb ballistic current as a major BPVE mechanism has not previously been made, as it requires a careful partition of separate current components. 

In rt-TDDFT \cite{ andrade-inq-2021, kononov-electron-2022, xu-real-time-2024}, the current operator in the velocity gauge is represented by its matrix elements
\begin{align}\label{eq:currop}
    \xi_{mm'}(\vec{k},t) & = \int d^3r \, \psi^{*}_{m\vec{k}}(\vec{r},t) \left ( \vec{\nabla}-i\vec{A}(t) + [V_\text{NL}, \vec{r}] \right ) \psi_{m'\vec{k}}(\vec{r},t)
\end{align}
for instantaneous (\(\psi\)) orbitals , where \(\vec{A}(t)\) is the uniform vector potential and \(V_\text{NL}\) is the nonlocal part of the Hamiltonian introduced by the pseudopotentials. We take \(\psi^0\) to represent the ground state Kohn-Sham orbitals. The total current of an \(N\)-electron system is then given by
\begin{align}\label{eq:current}
    \vec{J}(t) & = \mathrm{Im} \sum_{\vec{k}} \mathrm{Tr} \left [ \rho \, \xi(\vec{k},t) \right ] 
\end{align}
where $\rho$ is the density matrix. In the basis of instantaneous orbitals \(\psi\), it has nonzero elements \(\rho_{nn} = f_n = 1\) for \(n\le N\) and \(0\) elsewhere. 

The ballistic photocurrent is identified with the band diagonal contributions to the photocurrent, and can be written as a product of projected occupation numbers with band velocities,
\begin{equation}\label{eq:ball-current}
\vec{J}^\text{b}(t) = \sum_{nk} \tilde{f}_{nk}(t) v_{nk}    
\end{equation}
in accordance with the historical perturbative definition of the ballistic current~\cite{belinicher-relation-1988,sturman-ballistic-2019}. Here, \(\tilde{f}_{nk}(t) = \sum_m f_m \vert \langle \psi_{mk}(t)\vert \psi_{nk} ^0\rangle\vert^2 \) are the instantaneous occupation numbers of the Kohn-Sham ground state bands, and $v_{nk}$ are their band velocities. On the other hand, shift current has been identified as being driven by coherences (band off-diagonal density matrix elements) using perturbation theory arguments~\cite{tan-shift-2016, dai-recent-2023}. 

Previous attempts to separate band-diagonal and band-offdiagonal contributions in real-time simulations of the photocurrent have been made in the literature~\cite{he-ultrafast-2022}. However, we find that such approaches which rely on a transformation of Eq.~\ref{eq:current} to the basis of ground state Kohn-Sham bands are unsuitable here because of the slow convergence with the number of bands in the velocity gauge~\cite{virk-semiconductor-2007, ventura-gauge-2017}. Here, we instead take Eq.~\ref{eq:ball-current} as our definition of ballistic current.

To test the predictions of perturbation theory, we perform ab-initio rt-TDDFT simulations of photocurrents in monolayer GeS. This is a two-dimensional polar material which has previously been identified as a candidate with strong in-plane shift currents~\cite{rangel2017, chan-giant-2021,kushnir-ultrafast-2017,kushnir-ultrafast-2019}.  With the out-of-plane normal as the \(x\)-direction, it has a mirror plane in the \(y\)-direction and a strong ground state polarization in the \(z\)-direction. Therefore, its photocurrent is expected to be the strongest in the \(z\)-direction. 

In the following, we consider the longitudinal (\(z\)) photocurrent under monochromatic light linearly polarized in the \(z\)-direction, using resonant excitation frequencies from 2.4 eV to 3.2 eV, above the DFT band gap of 1.64 eV. 
While including electronic correlations at a hybrid functional or GW level increases the size of the band gap and changes it to a slightly indirect one~\cite{malone-quasiparticle-2013, li-germanium-2016}, our simulations at frequencies high above the band gap should be insensitive to the details of the band structure near the band edges. 
The continuous wave excitation is smoothly turned on at \(t = 0\) fs by convolving with a hyperbolic tangent function with a width of \(\eta=\) 0.1 fs, in order to avoid discontinuities in the dynamics. The resulting dynamics is insensitive to \(\eta\) in the range 0.1-1.0 fs. rt-TDDFT is performed with the PBE exchange-correlation functional, using norm conserving pseudopotentials, in the planewave basis at a kinetic energy cutoff of 50 Ha, and with a time step of 0.5 attosecond, as implemented in the INQ code~\cite{andrade-inq-2021,INQ}. Using a supercell geometry, the ion positions are fixed and the lattice constants are kept constant at \((a_x, a_y, a_z) = (15.0, 3.603, 4.523) \)\AA~throughout the trajectory. A \(k\)-point grid of 2$\times$8$\times$32 is sufficient for computing photocurrent under these conditions. 

Due to computational limitations, the  {sum in Eq.~\ref{eq:ball-current}} requires a finite cutoff on the number of ground state Kohn-Sham orbitals. For this, we used a total of \(M = 40\), comprising 20 occupied bands and 20 unoccupied bands.  {The results} are insensitive to further increase of \(M\), indicating that electronic populations  from higher lying bands are unimportant.

\begin{figure}[htbp!]
\centering
\includegraphics[width=0.45\textwidth]{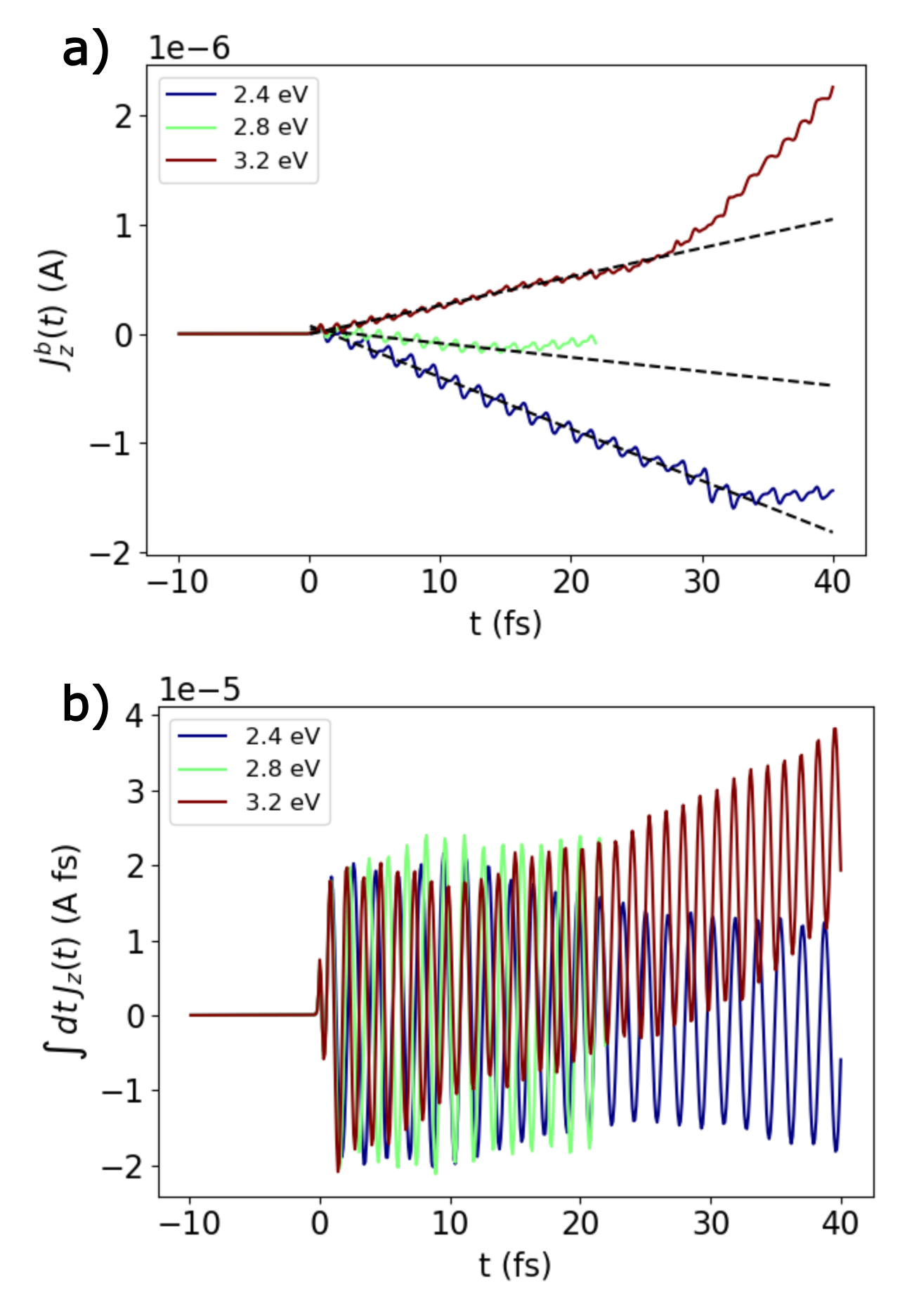}
\caption{\label{fig:current-time} a)  {Ballistic} photocurrent time-dependence of monolayer GeS under monochromatic continuous wave excitation with varying frequency and electric field strength  {0.1} V/nm, turned on at \(t = 0\). {Dashed black lines indicate linear fit at early times.  } b)  {Total} photocurrent under the same excitation conditions, displayed as a time-integral to illustrate changes in the direct current (dc) component. A linear rise in the time-integrated current indicates a  {constant} current state. }
\end{figure}

Over  {the initial 0 - 10 fs of the } trajectory, we find that  {ballistic} current first increases linearly for all excitation energies. The duration,  {magnitude, and sign} of this initial rise and the behavior after depends on the excitation energy. Then, the current starts {deviating from a linear time dependence} (Fig.~\ref{fig:current-time}a).  
According to perturbation theory, ballistic current is expected to increase linearly with time at short times~\cite{sipe-second-order-2000}, with the steady state current given by \(\tau\, dJ^\text{b}/dt\) for some relaxation time scale \(\tau\). {This departure from linearity reflects a true departure from perturbation theory dynamics, and is not a numerical artifact (Supplemental Material). Based on linear fits to the initial part of the trajectory (Fig.~\ref{fig:current-time}a), we estimate the relaxation times in monolayer GeS to be the time of deviation from linearity: $\tau=$ 30, 13, 22 fs, for 2.4 eV, 2.8 eV, and 3.2 eV excitation frequencies respectively}.
We attribute this relaxation to electron-charge density (e-d) Coulomb scattering because electron-phonon scattering is removed by fixing the ionic degrees of freedom in these simulations.
The saturation of ballistic current is not related to bleaching of the conduction/valence bands as it is observed even when carrier occupations are far from unity at low intensities (Supplemental Material). Besides the  {steadily growing} ballistic current, we also observe oscillations in \(J^\text{b}\) which increase with intensity. These are a result of higher order processes which will not be discussed further as they do not contribute to net current. {For the 2.8 eV excitation frequency, only data up to 22 fs is plotted because the dynamics of carrier scattering at this excitation frequency, beyond 22 fs, requires fine $k$-point sampling beyond current computational capabilities (see supplemental materials for full data set).}

 {The total current \(J(t)\)} has a different time-dependence (Fig.~\ref{fig:current-time}b). Its leading contributions are oscillations at the driving frequency, corresponding to direct absorption. 
 {Beyond that, there are direct current (zero-frequency) components, as seen from the steady rise in the time-integrated current over many periods of the driving frequency. For both the time-integrated total current and ballistic current, their change in sign across the frequency range 2.4 eV - 3.2 eV and their magnitudes and time dependence are comparable (Supplemental Material), indicating that ballistic current is strong in this system. Other zero-frequency contributions to the total current as computed in Eq.~\ref{eq:current} would include the shift current and anomalous velocity contributions~\cite{fregoso-bulk-2019}.  }
At certain frequencies, such as  {2.4 eV}, the system transitions from a state with low  {direct} current at short times to a state with larger  {direct} current. This crossover  {happens at comparable time scales} to the ballistic current relaxation time. 

\begin{figure*}[htbp!]
\centering
\includegraphics[width=\textwidth]{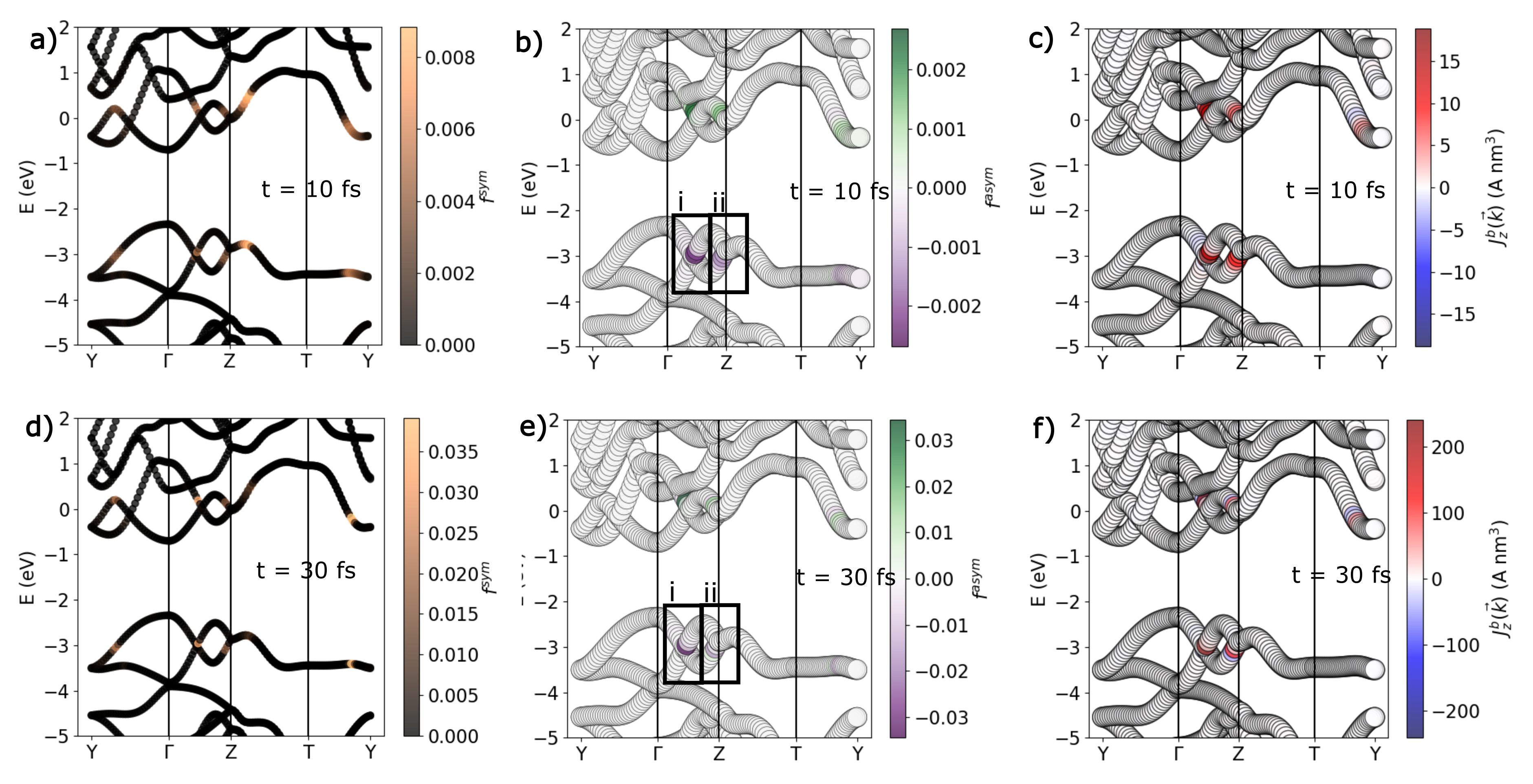}
\caption{\label{fig:bands} Band structure of monolayer GeS with instantaneous  {symmetric carrier occupation number,  antisymmetric} carrier occupation number, and  {ballistic} photocurrent superimposed, for \(t = 10~\mathrm{fs}\) after the start of continuous wave photoexcitation (a-c), and \(t = 30~\mathrm{fs}\) after the start of continuous wave photoexcitation (d-f).  Monochromatic excitation is at frequency 3.2 eV and electric field strength  {0.1} V/nm.  {Inset in (d) denotes the path \(\Gamma\) - \(Z\) - \(T\) - \(Y\) along which the band structure is plotted. Insets i, ii in (b,e) denotes two regions of carrier accumulation displaying different dynamics.} }
\end{figure*}

To understand these phenomena, we examine contributions to \(J^\text{b}\) across \(k\)-points and bands. Partitioning the  {ballistic} current as \(\vec{J}^\text{b}(t) \equiv \sum_{\vec{k};\,m=1}^{M} j_{m}(\vec{k})\), in Fig.~\ref{fig:bands} we plot the symmetrized ballistic current contribution \(j^\text{b}_m(k,t) = (j_{m}(k,t) + j_{m}(-k,t))/2\), alongside instantaneous carrier occupations. The antisymmetrized currents are not considered as their total does not contribute any current.  {Carrier occupation numbers are further separated into symmetrized \(\tilde{f}^{\text{sym}}_{nk} = (\tilde{f}_{nk} + \tilde{f}_{n,-k} )/2\) and antisymmetrized parts \(\tilde{f}^{\text{asym}}_{nk} = \tilde{f}_{nk} - \tilde{f}_{n,-k}\). }

By comparing carrier occupations at \(t = 10~\mathrm{fs}\) and \(t = 30~\mathrm{fs}\), we see that carriers are initially excited via resonant transitions ( {\(Z\)-\(T\) line in Fig.~\ref{fig:bands}a}), before relaxing via e-d scattering to isolated locations in the Brillouin zone ( {\(T\)-\(Y\) line in Fig.~\ref{fig:bands}a}).
These accumulation points tend to be located near van Hove singularities because of limited phase space for scattering there, but do not necessarily need to be at the valence band maximum or the conduction band minimum. When other relaxation processes are included, such as electron-phonon interactions, and at longer times, it is expected that there will be a greater tendency for relaxation to energy band extrema.  {The results shown Fig.~\ref{fig:current-time} are expected to be modified by these additional relaxation processes at times larger than the shortest relaxation time \(\tau\). Close to the recombination time, the presence of indirect band gaps may become important in stabilizing the antisymmetric and symmetric carrier occupation numbers.} Nevertheless, these results show that electron-density scattering drives transient carrier dynamics to band locations which affect ballistic currents. 

We find that ballistic current is mostly concentrated where carrier occupations are high, and where there is sufficient carrier velocity, such as along the \(\Gamma\)-\(Z\) and \(T\)-\(Y\) lines. This is consistent with the usual perturbative interpretation of ballistic current. In this picture, it is the antisymmetrized carrier occupations \(\tilde{f}^{\text{asym}}_{nk}\) that carry photocurrent, due to the time-reversal symmetry of band velocities \(v_{n,k} = -v_{n, -k}\).  {Accordingly, we find that ballistic current closely tracks the locations of antisymmetrized carriers in our simulations (comparing Fig.~\ref{fig:bands}b with Fig.~\ref{fig:bands}c and Fig.~\ref{fig:bands}e with Fig.~\ref{fig:bands}f). }  Despite the vicinity of the accumulation points  {(Fig.~\ref{fig:bands}d)} having large carrier densities,  {ballistic currents are not confined to those locations,} because the antisymmetrized carrier occupations  {do not follow the same dynamics as symmetrized carrier occupations.}

 {The crossover from low ballistic current at short times (\(t<\tau\)) to higher ballistic current at longer times is explained by the dynamics of the antisymmetrized carriers. For (\(t<\tau\)), antisymmetrized carriers are excited into bands resonant with the driving frequency, appearing at two locations along the \(\Gamma-Z\) line (i and ii in Fig.~\ref{fig:bands}b) {with negative signed contributions. After relaxation,  antisymmetrized carriers at location ii start to appear with both positive and negative contributions, while those at location i are unchanged in sign} (Fig.~\ref{fig:bands}e), driving the change in ballistic current magnitude before and after the relaxation time.  }

\begin{figure}[htbp!]
\centering
\includegraphics[width=0.45\textwidth]{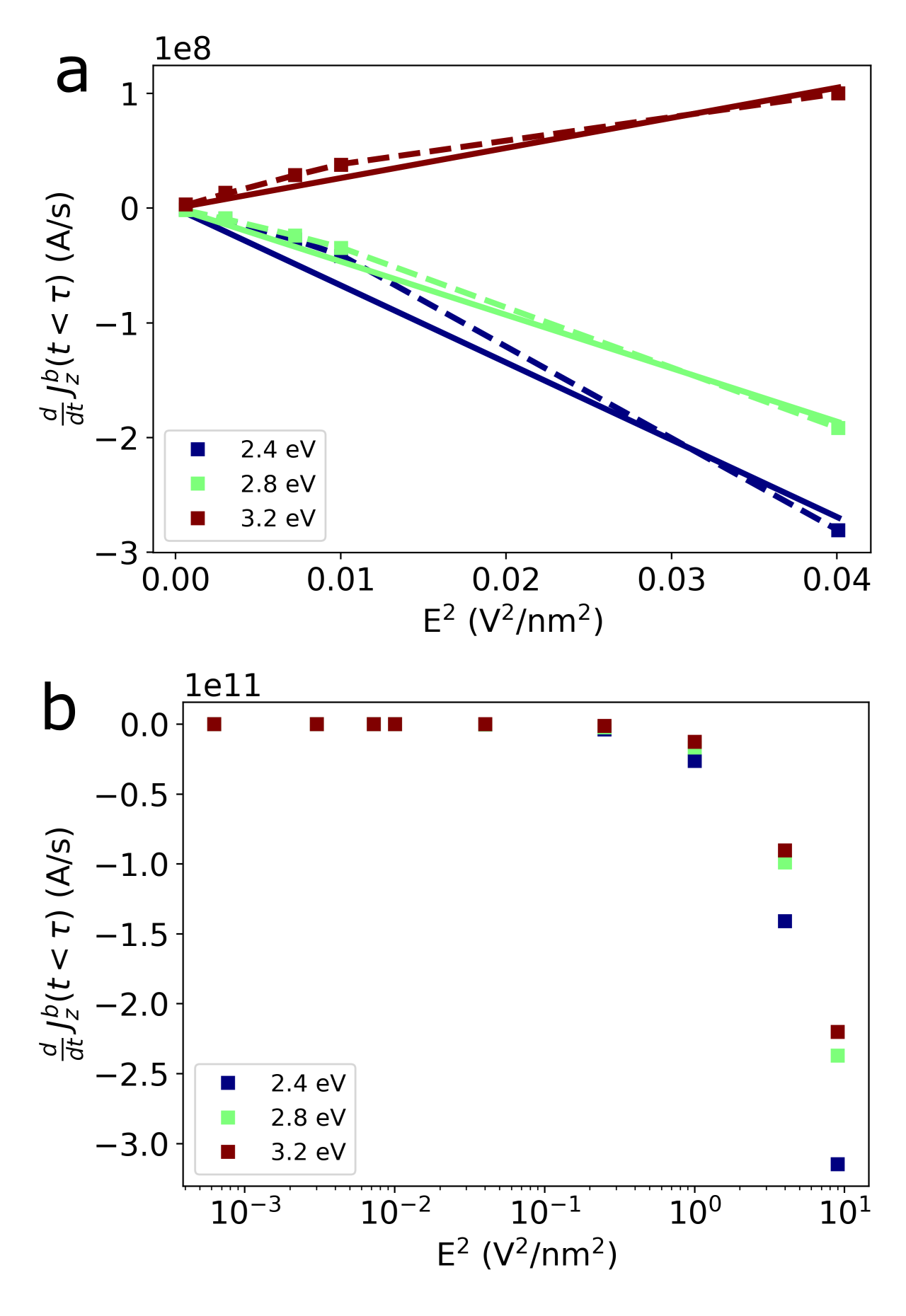}
\caption{\label{fig:power-freq}  Rate of change of  {ballistic} photocurrent at early times showing scaling with light intensity, for varying excitation energies.   { a) Data collected across the entire intensity range shows the onset of strong field behavior at around \(E\sim 1 \)V/nm. b) Low field data. } Colored lines are linear fits of data points,  {dashed lines are quadratic fits of data (including linear terms)} .}
\end{figure}

This  {ballistic current} picture is further supported by the intensity dependence of the measured currents.  {At low applied fields,} the rate of increase of ballistic currents before  {relaxation} (\(t<\tau\)) is  {almost} linear with intensity \(E^2\) (Fig.~\ref{fig:power-freq}a), agreeing with perturbation theory~\cite{sipe-second-order-2000}.   {On fitting the intensity dependence over this range as \(\frac{d}{dt}J_z^b(t<\tau) = A E^2+BE^4\) (Supplemental Material), we find a small \(E^4\) contribution, which is likely due to the system beginning to transition to a high-field regime at \(E=1\,\mathrm{V/nm}\). The high-field regime is associated with a large rise in the magnitude of the current (Fig.~\ref{fig:power-freq}b), and carrier occupation numbers approaching unity (Supplemental Material), indicating a breakdown of the perturbative picture.}

The optical responsivity of the ballistic current, \(\sigma = J/(b E^2)\), is estimated by taking the ballistic current at the  {relaxation} time (\(\tau\)), with \(b = a_x a_y\) being the cross-sectional area  {and $E$ the self-consistent electric field in the simulation. We consider only the current prior to the fastest relaxation time in this estimate because other relaxation processes not included in the present simulation may further modify the current after this time.}  
{At the excitation frequency of 3.2 eV, over  the relaxation time,} the maximum magnitude of ballistic current rises to {5.7}\(\times 10^{-7}~\mathrm{A}\)(Fig.~\ref{fig:current-time}). we estimate the ballistic current optical responsivity to be { 
\(1.1\times 10^{-4}~\mathrm{A/V^2}\)}, to be compared with the shift current optical responsivity of \(\sim 1\times 10^{-4}~\mathrm{A/V^2}\) obtained from perturbation theory~\cite{rangel2017}.  {Ballistic currents in the same material arising from the circular photogalvanic effect are of a similar order of magnitude. Using our simulated relaxation time of {13 fs for 2.8  eV excitation}, together with the response function computed in Ref.~\cite{panday-injection-2019}, we estimate a maximum response of {1.3}\(\times 10^{-4}~\mathrm{A/V^2}\) for the circular photogalvanic effect. }

{These large ballistic currents are a result of the two-dimensional nature of this material. The reduced screening of electron-density interactions in low dimensions leads to faster carrier scattering rates~\cite{latini-excitons-2015,ugeda-giant-2014}, increasing the likelihood of creating highly asymmetric carrier distributions. The reduced dimensionality may also constrain the number of scattering pathways, and enhance the joint density of states~\cite{cook-design-2017,fregoso-quantitative-2017}. As an illustration of this principle, we compute the ballistic current of {cubic} BN, a noncentrosymmetric three-dimensional semiconductor (Supplementary Material), finding that its ballistic current is negligible at comparable intensities. Atomically thin, polar materials would then be an ideal system for testing these predictions experimentally. } {On the other hand, strong many-body effects in low-dimensional materials may strongly modify the absorption edge, necessitating the use of advanced exchange-correlation energy functionals for treating excitations close to the absorption edge in rt-TDDFT~\cite{gruning-second-2014}. }

In summary, we have observed ballistic currents generated by Coulomb scattering in rt-TDDFT simulations. These currents are comparable in magnitude to shift currents. 
These observations suggest that a multitude of mechanisms contribute to the BPVE, with consequences for understanding many experiments beyond the narrow interpretation of the excitation shift current.   The ballistic current optical responsivity from Coulomb scattering measured here is similar to that from electron-phonon scattering ~\cite{dai-first-principles-2021}, which is expected considering that both forms of scattering have similar overall scattering rates~\cite{bernardi-abinitio-2014}, and suggesting that other forms of scattering, such as from defects or interfaces, could potentially contribute strongly to ballistic photocurrents as well. While we have shown that the Coulomb ballistic current is already present at the mean-field approximation, the effects of electron-electron correlation also warrants further study.  {Photo-Hall measurements and ultrafast time-resolved experiments will be able to distinguish the Coulomb ballistic current from other current contributions, by measuring the time scales of the rise of photocurrents, and the dynamics associated with the transient carrier accumulation points. }

\begin{acknowledgments}
Theory and simulation were supported by the Computational Materials Sciences Program funded by the US Department of Energy, Office of Science, Basic Energy Sciences, Materials Sciences and Engineering Division.  Additional data analysis was supported by a user project at the Molecular Foundry was supported by the Office of Science, Office of Basic Energy Sciences, of the U.S. Department of Energy under Contract No. DE-AC02-05CH11231. This research used resources of the National Energy Research Scientific Computing Center, a DOE Office of Science User Facility supported by the Office of Science of the U.S. Department of Energy under Contract No. DE-AC02-05CH11231.
AAC, XA and TA work was performed under the auspices of the US Department of Energy by Lawrence Livermore National Laboratory under contract DE-AC52-07NA27344.
\end{acknowledgments}

\bibliography{main}

\end{document}